\documentclass[journal=jacsat,manuscript=article]{achemso}

\usepackage{chemformula} 
\usepackage[T1]{fontenc} 
\usepackage{bm}
\usepackage[euler]{textgreek}
\usepackage{amssymb}
\usepackage{amsmath}
\usepackage{booktabs}
\usepackage{siunitx}
\usepackage{tikz}
\usepackage{setspace}
\usepackage{mathtools}

\author{Yeji Kim}
\author{Yoonho Jeong}
\author{Jihoo Kim}
\author{Eok Kyun Lee}
\affiliation[KAIST]
{Department of Chemistry, KAIST, Daejeon 34141, Korea}
\author{Won June Kim}
\affiliation[CNU]
{Department of Biology and Chemistry, Changwon National University, \\ Changwon 51140, Korea}
\author{Insung S. Choi}
\email{ischoi@kaist.ac.kr}
\affiliation[KAIST]
{Department of Chemistry, KAIST, Daejeon 34141, Korea}

\title
  {\Large MolNet: A Chemically Intuitive Graph Neural Network for Prediction of Molecular Properties}

\abbreviations{}

\begin{document}

\begin{abstract}

\begin{spacing}{1.2}
The graph neural network (GNN) has been a powerful deep-learning tool in chemistry domain, due to its close connection with molecular graphs. Most GNN models collect and update atom and molecule features from the fed atom (and, in some cases, bond) features, which are basically based on the two-dimensional (2D) graph representation of 3D molecules. Correspondingly, the adjacency matrix, containing the information on covalent bonds, or equivalent data structures (e.g., lists) have been the main core in the feature-updating processes, such as graph convolution. However, the 2D-based models do not faithfully represent 3D molecules and their physicochemical properties, exemplified by the overlooked field effect that is a “through-space” effect, not a “through-bond” effect. The GNN model proposed herein, denoted as MolNet, is chemically intuitive, accommodating the 3D non-bond information in a molecule, with a noncovalent adjacency matrix $\mathbf{\bar{A}}$, and also bond-strength information from a weighted bond matrix $\mathbf{B}$. The noncovalent atoms, not directly bonded to a given atom in a molecule, are identified within 5 Å of cut-off range for the construction of $\mathbf{\bar{A}}$, and $\mathbf{B}$ has edge weights of 1, 1.5, 2, and 3 for single, aromatic, double, and triple bonds, respectively. Comparative studies show that MolNet outperforms various baseline GNN models and gives a state-of-the-art performance in the classification task of BACE dataset and regression task of ESOL dataset. This work suggests a future direction of deep-learning chemistry in the construction of deep-learning models that are chemically intuitive and comparable with the existing chemistry concepts and tools.
\end{spacing}

\end{abstract}
\pagebreak

\section{Introduction}
The graph neural network (GNN) has recently become a powerful deep-learning (DL) model in chemistry, especially in the task of molecular properties and interactions. GNNs have intensively been used for regression tasks of molecular properties, such as solubility, lipophilicity, permeability, and atomization energy,\cite{1yang2019,2tang2020,3jo2020,4klicpera2020} and classification tasks of drug-target interactions. \cite{5torng2019,6nguyen2021,7jiang2020,8zhao2021,9wang2021} For example, the directed message passing neural network (D-MPNN) identified a new antibiotic compound, which is structurally distinct from conventional antibiotics, against a wide spectrum of pathogenic bacteria including \textit{Mycobacterium tuberculosis}, \textit{Clostridioides difficile}, and \textit{Acinetobacter baumanni}.\cite{10stokes2020}

Given a molecular graph composed of atoms as nodes and bonds as edges, the GNN models typically garner the atom-connectivity information (i.e., information on covalent bonds) via only adjacency matrix $\mathbf{A}$. However, these models do not faithfully take the three-dimensional (3D) molecular structures as input for the task from the chemist’s point of view. In the models, all the conformers of a molecule are represented as one simple undirected graph, losing the information on the relative positions of atoms in the space, and the conformers become indistinguishable from each other. 
In other words, a dynamic description of molecular properties and interactions is not considered. \cite{11schneider2010} Moreover, the 3D positions of atoms in a certain molecule (e.g., atom proximity in the 3D space) are equally crucial in the DL prediction of molecular properties. For example, the field effect (“through-space effect”)\cite{12golden1972,13goldberg1983,14hansch1991} states that the relative position of and the Euclidean distance between the non-bonded atoms of a molecule profoundly influence molecular properties, such as acidity. A famous textbook example of the field effect is the \textit{p}K difference (6.07 vs. 5.67 in 50\% aqueous ethanol (v/v) at 25 \textcelsius) between the \textit{syn} and \textit{anti} isomers of \textit{cis}-11,12-dichloro-9,10-dihydro-9,10-ethanoanthracene-2-carboxylic acid,\cite{15grubbs1971} which is not reliably manifested in the GNN models. Although there have been some attempts to represent 3D molecules in the GNN-based DL models, such as 3D graph convolutional network (3DGCN)\cite{16cho2019,17zhong2021} and spatial graph convolutional network (SGCN),\cite{18danel2020} these models only consider the 3D bond information $\bm{r}_{ij}=\bm{p}_{j}-\bm{p}_{i}$ ($\bm{p}_{i}$: position vector of atom $i$), not the relative positions of the non-bonded atoms in a molecule. In addition, not only the matrix $\mathbf{A}$ does not have the information on noncovalent atom locations in the 3D space, but it loses the information of bond type and strength, presented, for instance, as single and double straight lines in molecular graphs. Although the bond characteristics could be presented as edge features in GNN models, there have been no attempts to use a weighted adjacency matrix as input, which would be natural to the (organic) chemists.

In this work, we propose a GNN variant, denoted as MolNet, which contains chemically intuitive building blocks in its architecture. Specifically, we defined a noncovalent adjacency matrix $\mathbf{\bar{A}}$ and a weighted bond matrix $\mathbf{B}$ for the construction of MolNet, and investigated the model performance in the classification task of protein-ligand interactions.

\section{Results and discussion}

\begin{figure}[b!]
\centering
\includegraphics[width=0.5\linewidth]{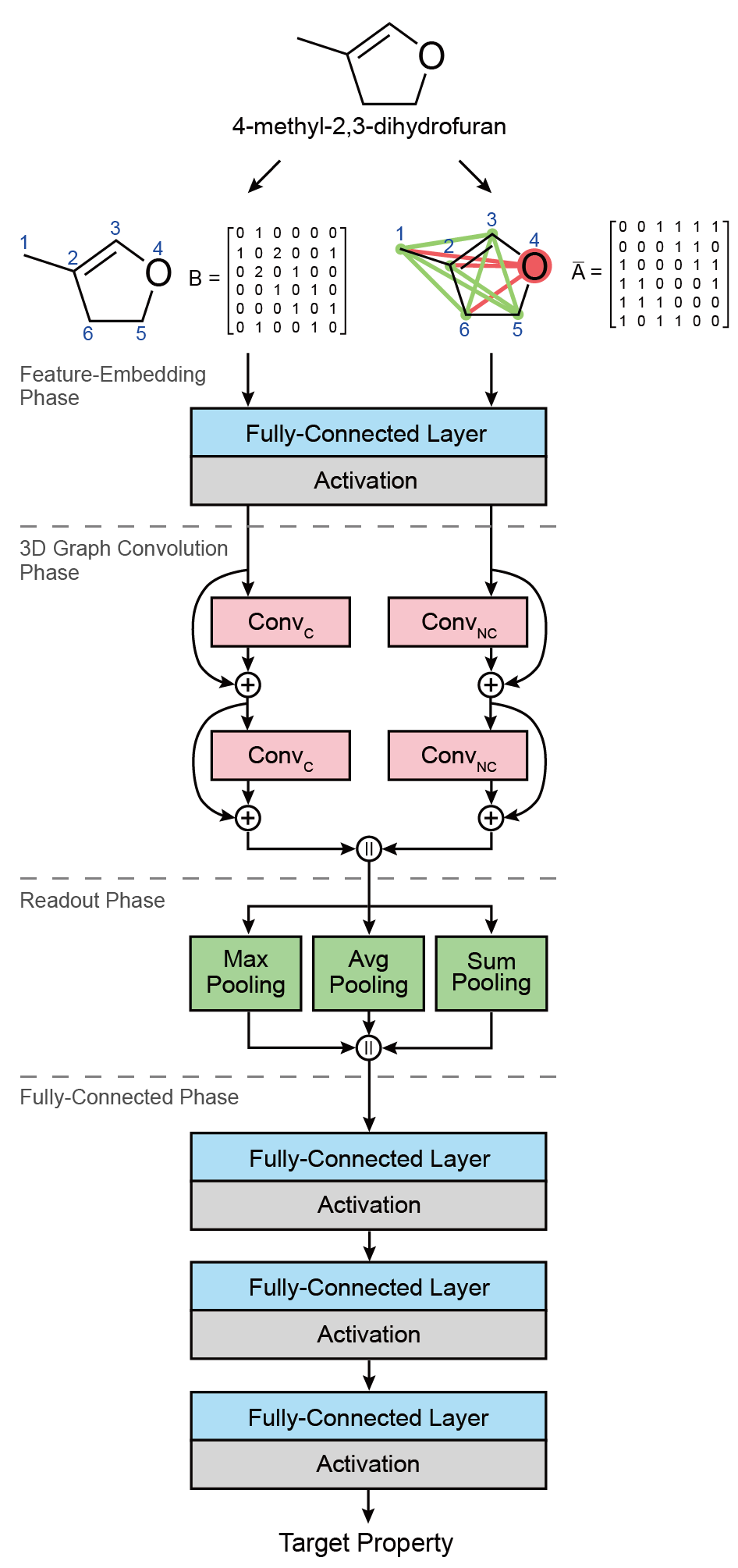}
\caption{(Top) The noncovalent adjacency matrix $\mathbf{\bar{A}}$ and weighted bond matrix $\mathbf{B}$ with 4-methyl-2,3-dihydrofuran as an example. Noncovalent interactions are indicated in green for C···C and in red for C···O. (Bottom) The MolNet architecture: feature-embedding, 3D graph convolution, readout, and fully-connected phases. $\oplus$: addition; $\parallel$: concatenation.}   
\label{fig:molnet_architecture}
\end{figure}

Figure~\ref{fig:molnet_architecture} shows the moduli introduced in this work, with 4-methyl-2,3-dihydrofuran as an example. The noncovalent adjacency matrix ($\mathbf{\bar{A}}\in\mathbb{R}^{N\times N}$, $N$: the number of atoms) contains the information on the relative positions of the non-bonded atoms in a molecule, which is used, complementary to covalent adjacency matrix ($\mathbf{A}$), in MolNet. Each element of the matrix $\mathbf{\bar{A}}$ is assigned as

\begin{singlespace}
\begin{equation}
   \mathrm{\bar{A}}_{ij}=\begin{cases}
   \ 1 & \quad \text{for noncovalent atom pairs} \\
   \ 0 & \quad \text{otherwise}
   \end{cases}
\label{eqn:noncovalent_adj}
\end{equation}
\end{singlespace}

\vspace{0.5\baselineskip} \noindent
where the noncovalent atom pairs, which are not directly bonded to each other, are the ones within an assigned cut-off distance.

The matrix $\mathbf{\bar{A}}$ could be considered as an intramolecular version of intermolecular (i.e., protein-ligand) connectivity matrices previously employed in PotentialNet\cite{19feinberg2018} and InteractionNet.\cite{20cho2020} Because the conventional adjacency matrix $\mathbf{A}$ does not fully convey the atomic-bond information for GNN training, we also define a weighted bond matrix $\mathbf{B}$, in which the edge weight is 1 for single bonds, 2 for double bonds, and 3 for triple bonds. The value of 1.5 is used as an edge weight for aromatic systems, such as benzene. Other resonance structures are not considered for the construction of $\mathbf{B}$.

\begin{figure}[b!]
\centering
\includegraphics[width=0.4\linewidth]{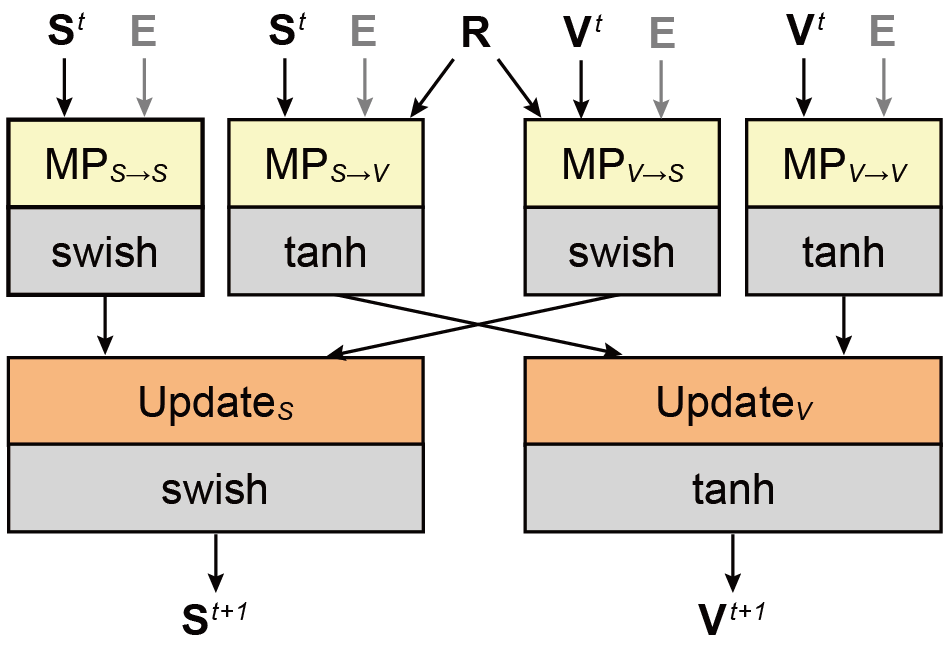}
\caption{3D graph convolution block in MolNet. $\mathbf{S}^{t}$ and $\mathbf{V}^{t}$ are scalar and vector features on the layer $t$, respectively. $\mathbf{R}$: relative atom position; $\mathbf{E}$: optional bond features (gray).}
\label{fig:molnet_conv}
\end{figure}

The overall architecture of the MolNet model consists of four phases (Figure~\ref{fig:molnet_architecture}): feature-embedding, 3D graph convolution, readout, and fully-connected phases. In the feature-embedding phase, the scalar features are equal to general atom features, which are embedded into a fully-connected layer, and the initial vector features are set to be zeros. In the 3D graph convolution phase, two convolution blocks, Conv\textsubscript{C} and Conv\textsubscript{NC}, are stacked and skip-connected. The graph convolution of Conv\textsubscript{C} is defined with atom features, relative atom positions, and the matrix $\mathbf{B}$ (instead of $\mathbf{A}$), and that of Conv\textsubscript{NC} with atom features, relative atom positions, and the matrix $\mathbf{\bar{A}}$. In the readout phase, the concatenated scalar features are aggregated along the atom axis. We use the multi-pooling, a simple variant of the mixed\cite{21yu2014,22lee2016} and set2set pooling,\cite{23vinyals2015} in which the max-, sum-, and average (avg)-pooled features are concatenated. The pooled features are finally fed into a set of fully-connected layers for prediction.

The detailed structures of the 3D graph convolution phase in MolNet are shown in Figure 2. The Conv\textsubscript{C} and Conv\textsubscript{NC} are built with four layers independently: message-passing (MP), activation, update, and activation layers. The MP layer makes new features from scalar (\textit{s}) or vector (\textit{v}) atom-features (\textit{s}-to-\textit{s}, \textit{s}-to-\textit{v}, \textit{v}-to-\textit{s}, \textit{v}-to-\textit{v}). Bond features could be added for message passing as an option. The activation layer converts the input features to new non-linear output features, and swish and tanh are used for scalar and vector features, respectively. The activated features are aggregated into new scalar or vector atom features by the update layer, in which the scalar features are updated from \textit{s}-to-\textit{s} and \textit{v}-to-\textit{s} features, and the vector ones are from \textit{s}-to-\textit{v} and \textit{v}-to-\textit{v} features (See the supporting information (SI) for detailed mathematical expression). The skip connection\cite{24he2016} to each Conv\textsubscript{C} or Conv\textsubscript{NC} block is adopted in such a fashion that the input features of each block are kept and then added to the output features.

\subsection{MolNet performance on BACE dataset}

Model performance was investigated with the BACE dataset. The BACE dataset provides 1547 experimental binary inhibition labels with the 3D positioned ligands aligned to the binding pocket of human \textbeta-secretase 1 (BACE-1).\cite{25subramanian2016} BACE-1 is an aspartic acid protease, which is involved in the generation of amyloid-\textbeta (A\textbeta) in neurons, and has been one of the targets for the treatment of Alzheimer’s disease. We used 1478 ligands (653 active and 825 inactive ligands with an IC\textsubscript{50} threshold of 100 nM) after filtering out large ligands that contained more than 200 heavy atoms. The model performance was evaluated with the metrics commonly used in binary-classification problems: the area under the curve-receiver operating characteristic (AUC-ROC) and the area under the curve-precision-recall (AUC-PR) values.\cite{26davis2006,27ozenne2015} The ROC curve shows how the true positive rate (TPR) varies with the false positive rate (FPR), and the PR curve does how precision (TP divided by the sum of TP and FP) varies with TPR (a.k.a., recall or sensitivity). All the experiments were performed by using 10-fold cross validation.

\begin{figure}[t!]
\centering
\includegraphics[width=0.5\linewidth]{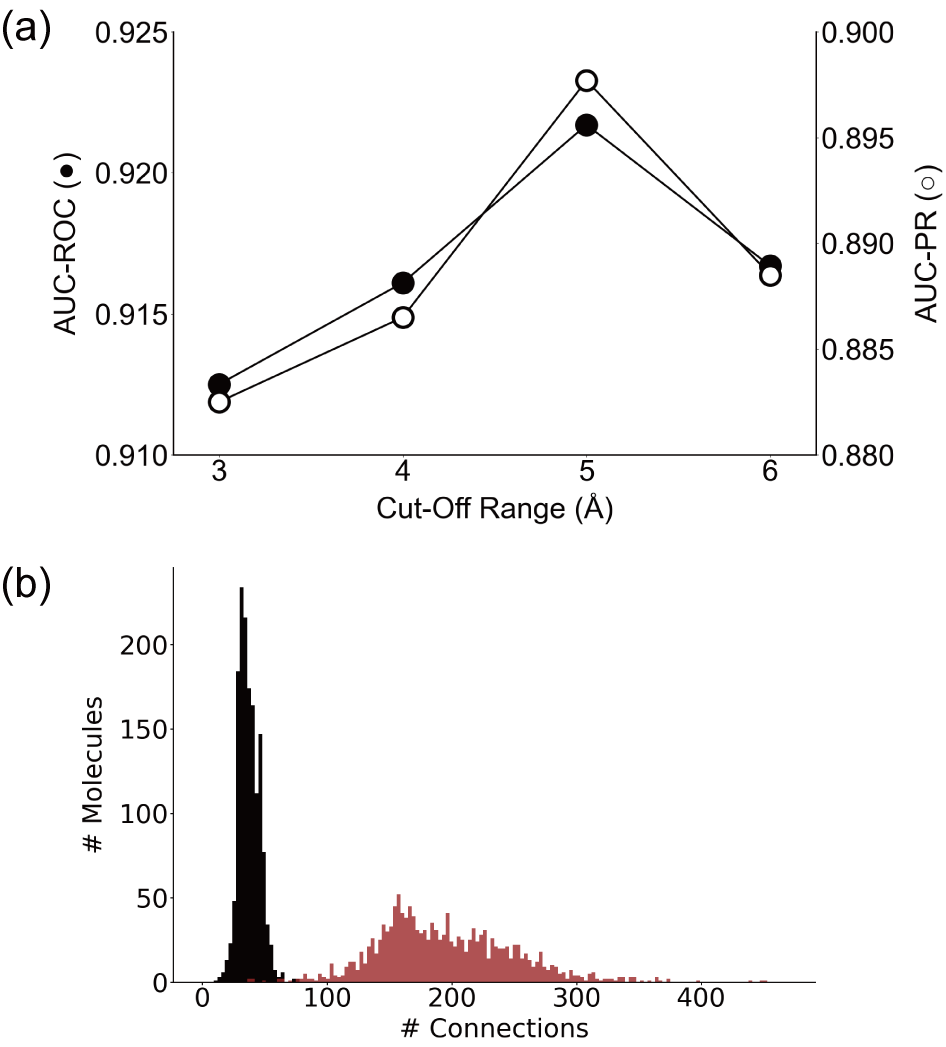}
\caption{BACE dataset analysis. (a) Graph of AUC-ROC and AUC-PR values of versus cut-off range. (b) Histograms of (black) the number of covalent bonds and (red) the number of noncovalent interactions within 5-Å cut-off range.}
\label{fig:bace_search}
\end{figure}

We first screened the cut-off ranges for $\mathbf{\bar{A}}$ from 3 to 6 Å with a 1-Å increment (Figure ~\ref{fig:bace_search}a). The model performance, evaluated with AUC-ROC and AUC-PR, was observed to be the highest with 5 Å of cut-off range and it decreased with a larger cut-off range, which was the same trend as the result for InteractionNet.\cite{20cho2020} For example, the AUC-ROC and AUC-PR values were 0.9217 and 0.8977 for 5-Å cut-off range, and 0.9167 and 0.8885 for 6-Å range. The enhanced performance with the increased number of noncovalent atom pairs (i.e., more ones in $\mathbf{\bar{A}}$) below 5 Å of cut-off range quantitatively supported the use of $\mathbf{\bar{A}}$ for molecular representations in MolNet. Based on the screening results, the optimal cut-off range for the $\mathbf{\bar{A}}$ construction was set to be 5 Å. Given the cut-off range of 5 Å, we further analyzed the distributions in the number of covalent bonds (for the construction of $\mathbf{B}$) and the number of non-bonded interactions (for the construction of $\mathbf{\bar{A}}$) for the BACE dataset (Figure~\ref{fig:bace_search}b). The distribution analysis showed the bond number of 36.85 ± 8.41 for $\mathbf{B}$ and the interaction number of 194.50 ± 54.72 for $\mathbf{\bar{A}}$ on average. In other words, the number of the non-bonded intramolecular interactions of the atoms in a BACE-1 ligand within a range of 5 Å were about 5 times more than that of the covalent bonds in the ligand, which would additionally rationalize the $\mathbf{\bar{A}}$ implementation to GNN models.

\begin{table}[t!]
  \caption{Ten-fold cross-validation results for MolNet compared with baseline models on the BACE dataset.}
  \label{tbl:bace}
  \begin{tabular}{c|cc}
  \hline
  Model Type      & AUC-ROC         & AUC-PR          \\
  \hline
  GCN             & 0.8713          & 0.8335          \\
  Weave           & 0.8763          & 0.8368          \\
  MPNN            & 0.8602          & 0.8012          \\
  3DGCN           & 0.8800          & 0.8371          \\
  \textbf{MolNet} & \textbf{0.9217} & \textbf{0.8977} \\
  \hline
  \end{tabular}
\end{table}

\begin{figure}[b!]
\centering
\includegraphics[width=0.6\linewidth]{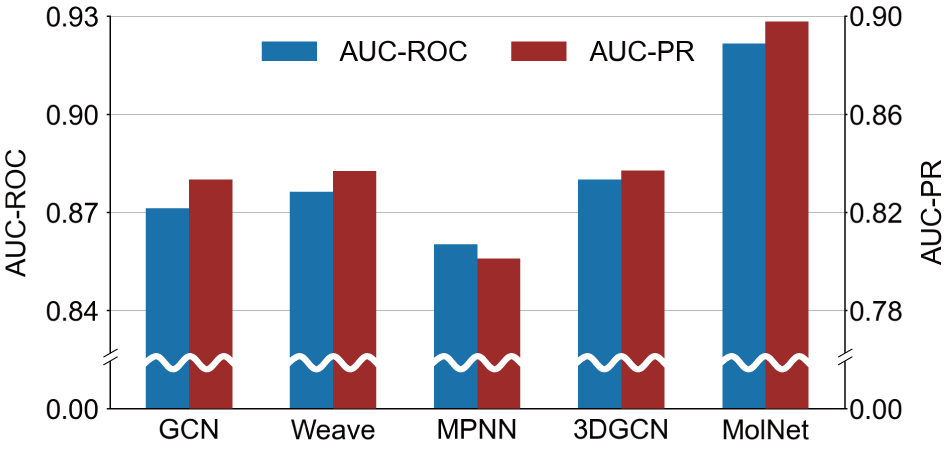}
\caption{Ten-fold cross-validation results for MolNet compared with baseline models on the BACE dataset.}
\label{fig:molnet_result}
\end{figure}

We compared the binary-classification performance of MolNet for the BACE task with various GNN models, such as GCN,\cite{28kipf2016} 3DGCN,\cite{16cho2019} Weave,\cite{29kearnes2016} and MPNN,\cite{30gilmer2017} as baseline models. The GCN is considered as a forerunner of spectral GNNs, in which graph convolution is conducted only with atom features via a normalized adjacency matrix with self-loop. 3DGCN is one of the 3D GNNs, which takes both atom features and 3D interatomic positions. The Weave model updates both atom and bond features in the convolution, and the MPNN updates atom features by aggregating the messages made from the embedded features of neighbor atoms and bonds. The performance-comparison results for the BACE dataset showed that MolNet significantly outperformed all the baseline models, showing a state-of-the-art performance (Table~\ref{tbl:bace} and Figure~\ref{fig:molnet_result}). The AUC-ROC and AUC-PR values for MolNet were 0.9217 and 0.8977, respectively, which were noticeably higher than the baseline models. For example, the 3DGCN model produced the best performance among the baseline models with 0.8800 of AUC-ROC and 0.8371 of AUC-PR, which were enhanced by 4.7\% and 7.2\% for MolNet. The results implied that the implementation of $\mathbf{\bar{A}}$ and $\mathbf{B}$ to MolNet made the model extract better chemical features for predicting the ligand interactions to BACE-1.

\begin{table}[]
  \caption{Comparative studies.}
  \label{tbl:comparative_studies}
  \begin{tabular}{c|cc}
  \hline
  Model Type      & AUC-ROC         & AUC-PR          \\
  \hline
  3DGCN\textsubscript{[A]}             & 0.8800          & 0.8371          \\
  MolNet\textsubscript{[A][+BF]}   & 0.9162          & 0.8807          \\
  MolNet\textsubscript{[A][$-$BF]}   & 0.9170          & 0.8863          \\
  3DGCN\textsubscript{[B]}             & 0.8980          & 0.8729          \\
  MolNet\textsubscript{[+BF]} & 0.9122 &  0.8871 \\
  \textbf{MolNet} & \textbf{0.9217} & \textbf{0.8977} \\
  \hline
  \end{tabular}
\end{table}

It is also to note that we did not use the bond features as input for the MolNet model, although it has commonly been observed that the addition of bond features generally enhances the model performance.\cite{31coley2017} In our case, the addition of bond features (e.g., aromaticity, bond chirality, bond conjugation, and ring structure) rather deteriorated the model performance, and the AUC-ROC and AUC-PR values decreased to 0.9122 and 0.8871, respectively (MolNet\textsubscript{[+BF]} in Table~\ref{tbl:comparative_studies}). Another control experiment was also performed, where $\mathbf{B}$ was changed back to $\mathbf{A}$ in MolNet, and the $\mathbf{A}$-implemented model was fed with or without the bond features (MolNet\textsubscript{[A][+BF]} and MolNet\textsubscript{[A][$-$BF]}). The change of $\mathbf{B}$ to $\mathbf{A}$ (i.e., MolNet\textsubscript{[A][$-$BF]}) decreased the model performance (AUC-ROC: 0.9170; AUC-PR: 0.8863), and the addition of the bond features further decreased the model performance slightly (AUC-ROC: 0.9162; AUC-PR: 0.8807). These comparison analyses confirmatively supported the use of weighted bond matrix $\mathbf{B}$ instead of $\mathbf{A}$, at least for our model that did not require the bond features for performance enhancement. The performance enhancement with $\mathbf{B}$ (3DGCN\textsubscript{[B]}) was also observed for 3DGCN: the AUC-ROC and AUC-PR values increased to 0.8980 and 0.8729, respectively, from 0.8800 and 0.8371. These results indicated that the MolNet model was well-adapted to the molecular systems with $\mathbf{\bar{A}}$ and $\mathbf{B}$, and the bond-related information might be superfluous at least in our case.

\begin{figure}[t!]
\centering
\includegraphics[width=0.8\linewidth]{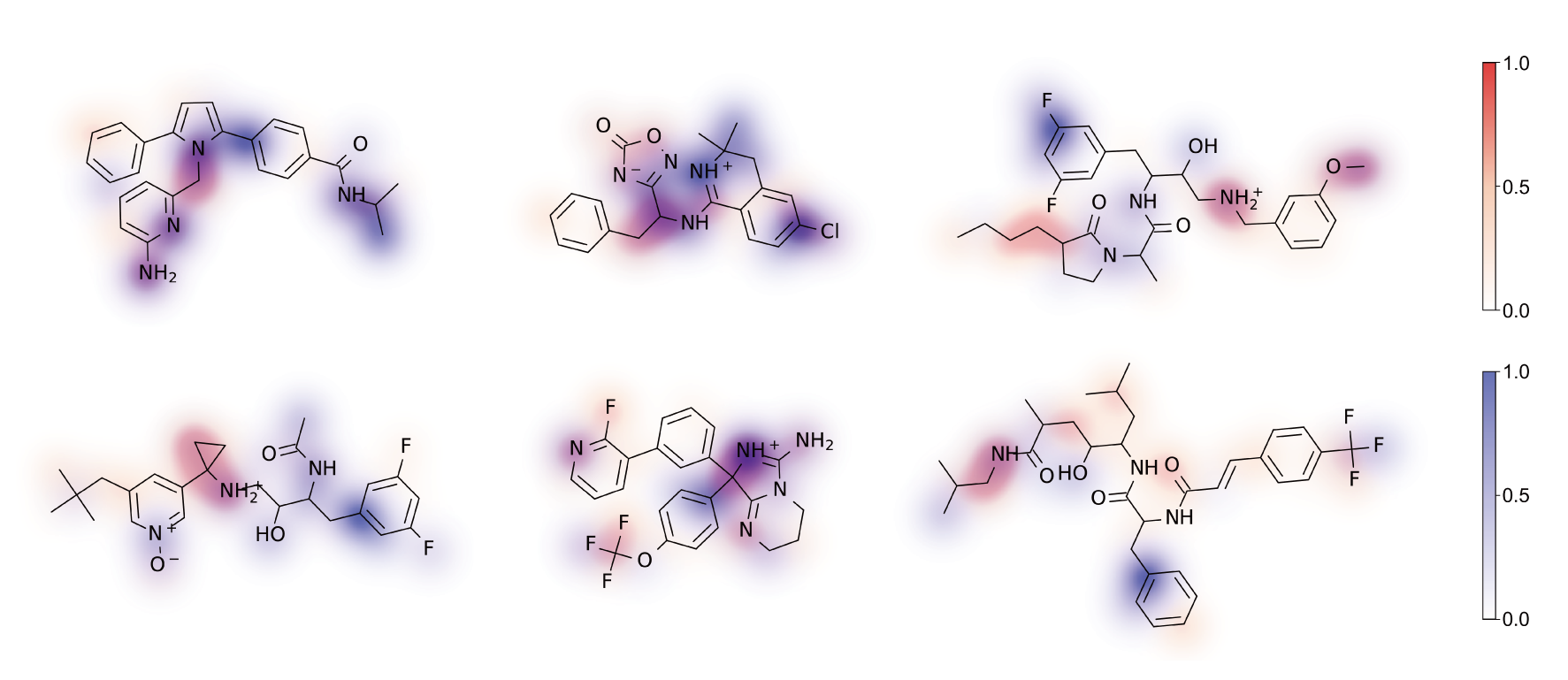}
\caption{Heat maps of atomic contributions to binding-affinity prediction. Atom-level bond (for $\mathbf{B}$) and non-bond (for $\mathbf{\bar{A}}$) contributions are represented in red and blue, respectively, with color intensity proportional to relative contribution magnitude.}
\label{fig:molnet_heatmap}
\end{figure}

MolNet supposedly learned the bond (e.g., inductive and resonance effects) and non-bond (e.g., field effect) contributions to molecular properties via $\mathbf{B}$ and $\mathbf{\bar{A}}$, respectively. Each contribution could be visualized in the heat map, in which the contributions were represented in red and blue, respectively (Figure~\ref{fig:molnet_heatmap}). Of significant note, all the ligands investigated showed different characteristics on each contribution for the BACE task. For example, the heat map indicated that, for the compound CHEMBL1092228 (top left in Figure~\ref{fig:molnet_heatmap}), the weighted bond matrix $\mathbf{B}$ emphasized on polar moieties (e.g., amino functional groups), while the noncovalent adjacency matrix $\mathbf{\bar{A}}$ focused more on hydrophobic interactions (e.g., nonpolar alkyl chains). Similar trends were observed for other compounds, although the rules were not strict: in the case of the compound CHEMBL1257533 (bottom left in Figure~\ref{fig:molnet_heatmap}), a high bond contribution was observed at the aminocyclopropyl group, with assistance from the \textit{tert}-butyl group, and a high non-bond one at the benzyl and acetyl groups. The heat-map analysis clearly showed that MolNet enhanced its molecular-property prediction by additionally collecting the interaction information between the atoms that are not directly bonded covalently, but proximate to each other in the 3D space, which had been overlooked in DL chemistry.

\subsection{MolNet performance on ESOL dataset}

\begin{table}
  \caption{Ten-fold cross-validation results for the ESOL dataset.}
  \label{tbl:esol}
  \begin{tabular}{c|cc}
  \hline
  Model Type      & AUC-ROC         & AUC-PR          \\
  \hline
  GCN             & 0.4779          & 0.6316          \\
  Weave           & 0.4818          & 0.6398          \\
  MPNN            & 0.5131          & 0.6880          \\
  3DGCN           & 0.4557          & 0.5948          \\
  \textbf{MolNet} & \textbf{0.4300} & \textbf{0.5693} \\
  \hline
  \end{tabular}
\end{table}

We additionally compared the model performance for the regression task of the ESOL dataset, which contains 1128 aqueous solubility values in log(mol·L\textsuperscript{-1}) with SMILES-encoded molecules.\cite{32delaney2004} For model training and testing, we generated the optimized 3D molecular structures for the ESOL dataset with the Merck molecular force field (MMFF94).\cite{33halgren1996,34ebejer2012} The optimal cut-off range search for $\mathbf{\bar{A}}$ from 3 to 6 Å (optimal cut-off range: 5 Å), and the distributions in the number of covalent bonds and the number of noncovalent interactions within 5-Å cut-off range are shown in the SI. The mean-absolute-error (MAE) and the root-mean-square-error (RMSE) were used as metrics for the regression task. Table~\ref{tbl:esol} shows that MolNet exhibited the best performance among the models tested (i.e., GCN, Weave, MPNN, 3DGCN, and MolNet), and the performance enhancement was, compared with 3DGCN that exhibited the best performance among the baseline models, 5.5\% or 4.3\% for MAE or RMSE, respectively. Taken all together, the results clearly showed that the chemically intuitive MolNet model had great potential in the DL-based prediction of molecular properties.

\section{Methods}

MolNet learns two 3D molecular graphs: covalent molecular graph $\mathcal{G}_C\left(\mathcal{V},\mathcal{E}\right)$ with the input of $\mathbf{\mathcal{B}}\in\mathbb{R}^{n\times n}$, $\mathbf{X}\in\mathbb{R}^{n\times m}$, $\mathbf{E}\in\mathbb{R}^{n\times n\times e}$, and $\mathbf{R}\in\mathbb{R}^{n\times n\times 3}$, and noncovalent molecular graph $\mathcal{G}_{NC}\left(\mathcal{V},\mathcal{E}\right)$ with the input of $\mathbf{\bar{\mathcal{A}}}\in\mathbb{R}^{n\times n}$, $\mathbf{X}\in\mathbb{R}^{n\times m}$, and $\mathbf{R}\in\mathbb{R}^{n\times n\times3}$. $\mathbf{\mathcal{B}}$ is the normalized covalent adjacency matrix with self-loop, defined as ${\hat{\mathbf{D}}}^{-\mathbf{1}/\mathbf{2}}\hat{\mathbf{B}}{\hat{\mathbf{D}}}^{-\mathbf{1}/\mathbf{2}}$ ($\hat{\mathbf{B}}=\mathbf{B}+\mathbf{I}$, $\mathbf{D}$: diagonal node-degree ($d$) matrix ($\mathbf{D}=diag(d)$), $\mathbf{I}$: identity matrix), and $\mathbf{\bar{\mathcal{A}}}$ is the normalized noncovalent adjacency matrix of $\mathbf{\bar{A}}$ with self-loop in the same way as $\mathbf{\mathcal{B}}$.  $\mathbf{X}$ is the atom-feature matrix ($n$: maximum number of atoms per molecule in the dataset, $m$: the size of total atom features), $\mathbf{E}$ is the bond-feature tensor ($n$: maximum number of bonds per molecule in the dataset, $e$: the size of total bond features), and $\mathbf{R}$ is the 3D interatom-position tensor ($\bm{r}_{ij}=\bm{p}_{j}-\bm{p}_{i}$, $\bm{p}_{i}$: 3D (\textit{x}, \textit{y}, and \textit{z})-coordinate vector of an atom \textit{i}). Atoms are represented by scalar features $\mathbf{S}_C\in\mathbb{R}^{n\times m}$ and vector features $\mathbf{V}_C\in\mathbb{R}^{n\times3\times m}$ in $\mathcal{G}_C\left(\mathcal{V},\mathcal{E}\right)$ and by scalar features $\mathbf{S}_{NC}\in\mathbb{R}^{n\times m}$ and vector features $\mathbf{V}_{NC}\in\mathbb{R}^{n\times3\times m}$ in $\mathcal{G}_{NC}\left(\mathcal{V},\mathcal{E}\right)$, where the initial scalar features are the input atom features (i.e., ${\mathbf{S}_{C}^{(0)}}\in\mathbb{R}^{n\times m}=\mathbf{X}$ and ${\mathbf{S}_{NC}^{(0)}}\in\mathbb{R}^{n\times m}=\mathbf{X}$), and the initial vector features are set to be all-zero tensor (${\mathbf{V}_{C}^{(0)}}\in\mathbb{R}^{n\times3\times m}=\mathbf{O}$ and ${\mathbf{V}_{NC}^{(0)}}\in\mathbb{R}^{n\times3\times m}=\mathbf{O}$). The 3D graph convolutions, Conv\textsubscript{C} and Conv\textsubscript{NC}, are performed independently, and the results are concatenated at the final stage. 

In the message-passing (MP) operation, a set of features ($\left\{\mathbf{S}_C,\mathbf{V}_C\right\}$ and $\left\{\mathbf{S}_{NC},\mathbf{V}_{NC}\right\}$) is updated by first mixing with the atom (and optional edge) features of neighbor atoms, leading to the generation of an intermediate atom feature. For example, the scalar-to-scalar ($s\rightarrow v$) mixing (MP$_{s\rightarrow v}$ in Figure~\ref{fig:molnet_conv}), produces $\bm{m[s\rightarrow v]}$ after nonlinear activation. The $s\rightarrow s$ and $v\rightarrow v$ mixings involve the linear combination of atom features, followed by nonlinear activation (swish for $s\rightarrow s$ and tanh for $v\rightarrow v$). The $v\rightarrow s$ interconversion is made, after linear combination of atom features of two nodes, by the multiplication of the \textit{k}-axis elements of a vector feature with ${r}_{ij,k}$ ($k$: one of the three Cartesian axes) and subsequent feature-wise summation. The $s\rightarrow v$ interconversion is done by the tensor product ($\otimes$) with $\bm{r}_{ij}$ (See the SI for detailed mathematical expressions). The swish and tanh functions are used for $v\rightarrow s$ and $s\rightarrow v$ interconversions, respectively. The graph update operation involves the aggregation of $\bm{m[s\rightarrow s]}$ and $\bm{m[v\rightarrow s]}$ for scalar features or of $\bm{m[v\rightarrow v]}$ and $\bm{m[s\rightarrow v]}$ for vector features:

\begin{flalign}
\bm{m[s\rightarrow s]}_{ij}^{\left(t\right)}=
\left(
    \mathbf{S}_{i}^{\left(t\right)} \parallel \mathbf{S}_{j}^{\left(t\right)} \right)
    {\mathbf{W}_{s\rightarrow s \vphantom{i}}^{\left(t\right)}}
    +{\mathbf{b}_{s\rightarrow s \vphantom{i}}^{\left(t\right)}} 
\label{eqn:m_stos}
\end{flalign}

\begin{flalign}
\bm{m[v\rightarrow v]}_{ij}^{\left(t\right)}=
\left(
    \mathbf{V}_{i}^{\left(t\right)} \parallel \mathbf{V}_{j}^{\left(t\right)} \right) 
    {\mathbf{W}_{v\rightarrow v \vphantom{i}}^{\left(t\right)}}
    +{\tilde{\mathbf{b}}_{v\rightarrow v \vphantom{i}}^{\left(t\right)}}
\label{eqn:m_vtov}
\end{flalign}

\begin{flalign}
\bm{m[v\rightarrow s]}_{ij}^{\left(t\right)}=
\sum_{k\in\left(x,y,z\right)} {
        \left(
            {
            \left (\mathbf{V}_{i}^{\left(t\right)} \parallel  \mathbf{V}_{j}^{\left(t\right)}\right)
            {\mathbf{W}_{v\rightarrow s \vphantom{i}}^{\left(t\right)}}
            +{\tilde{\mathbf{b}}_{v\rightarrow s \vphantom{i}}^{\left(t\right)}}}
            \right)}_{k} {r}_{ij,k} 
\label{eqn:m_vtos}
\end{flalign}

\begin{flalign}
\bm{m[s\rightarrow v]}_{ij}^{\left(t\right)}=\left(
    \left(
    \mathbf{S}_{i}^{\left(t\right)} \parallel  \mathbf{S}_{j}^{\left(t\right)}\right) 
    {\mathbf{W}_{s\rightarrow v \vphantom{i}}^{\left(t\right)}}
    +{\mathbf{b}_{s\rightarrow v \vphantom{i}}^{\left(t\right)}}
    \right) \otimes \bm{r}_{ij}
\label{eqn:m_stov}
\end{flalign}

\vspace{0.5\baselineskip} \noindent
where ${\mathbf{S}_{i}^{\left(t\right)}}$ and ${\mathbf{V}_{i}^{\left(t\right)}}$ are the scalar and vector features of the \textit{i}th atom on the layer \textit{t}, $\parallel$ represents the concatenation, $\mathbf{W}$ (
${\mathbf{W}_{s\rightarrow s\vphantom{i}}^{\left(t\right)}} \in\mathbb{R}^{{2m}\times {hp}_1}$, ${\mathbf{W}_{v\rightarrow v\vphantom{i}}^{\left(t\right)}} \in\mathbb{R}^{{2m}\times {hp}_1}$, ${\mathbf{W}_{v\rightarrow s\vphantom{i}}^{\left(t\right)}} \in\mathbb{R}^{{2m}\times {hp}_1}$, ${\mathbf{W}_{s\rightarrow v\vphantom{i}}^{\left(t\right)}} \in\mathbb{R}^{{2m}\times {hp}_1}$
) and $\mathbf{b}$ (
${\mathbf{b}_{s\rightarrow s\vphantom{i}}^{\left(t\right)}} \in\mathbb{R}^{{hp}_1}$,
${\mathbf{b}_{v\rightarrow v\vphantom{i}}^{\left(t\right)}} \in\mathbb{R}^{{hp}_1}$,
${\mathbf{b}_{v\rightarrow s\vphantom{i}}^{\left(t\right)}} \in\mathbb{R}^{{hp}_1}$,
${\mathbf{b}_{s\rightarrow v\vphantom{i}}^{\left(t\right)}} \in\mathbb{R}^{{hp}_1}$
) are the learnable weight matrices and biases, respectively. The ${hp}_1$ and ${hp}_2$ are the dimension sizes of output messages, and 
$\tilde{\mathbf{b}} \coloneqq \left[ \mathbf{b}\parallel\mathbf{b}\parallel\mathbf{b} \right]^{T}$ 
where $\tilde{\mathbf{b}}\in\mathbb{R}^{3\times {hp}_1}$. 
The same $\mathbf{W}$ and $\mathbf{b}$ are applied for all the graphs.

The graph update operation involves the aggregation of $\bm{m[s\rightarrow s]}$ and $\bm{m[v\rightarrow s]}$ for scalar features or of $\bm{m[v\rightarrow v]}$ and $\bm{m[s\rightarrow v]}$ for vector features:

\vspace{0.5\baselineskip}
\begin{flalign}
\mathbf{S}_{i}^{\left(t+1\right)} & =\mathbf{S}_{i}^{\left(t\right)}+\mathrm{swish}\left[\mathrm{Update}_{s}\right]  \nonumber \\
&   =\mathbf{S}_{i}^{\left(t\right)}+\mathrm{swish}\left[
\sum_{j\in\left(i,\mathcal{N}_i\right)}\mathrm{A}_{ij}\left(
\left(\bm{m[s\rightarrow s]}_{ij}^{\left(t\right)}\parallel \bm{m[v\rightarrow s]}_{ij}^{\left(t\right)} \right)
{\mathbf{W}_{s\vphantom{i}}^{\left(t\right)}}+{\mathbf{b}_{s\vphantom{i}}^{\left(t\right)}}
\right)\right]  
\label{eqn:update_s}
\end{flalign}

\begin{flalign}
\mathbf{V}_{i}^{\left(t+1\right)}   
&   =\mathbf{V}_{i}^{\left(t\right)}+\mathrm{swish}\left[\mathrm{Update}_{s}\right]  \nonumber \\
&   =\mathbf{V}_{i}^{\left(t\right)}+\mathrm{swish}\left[
\sum_{j\in\left(i,\mathcal{N}_i\right)}\mathrm{A}_{ij}\left(
\left(\bm{m[v\rightarrow v]}_{ij}^{\left(t\right)}\parallel \bm{m[s\rightarrow v]}_{ij}^{\left(t\right)} \right)
{\mathbf{W}_{v\vphantom{i}}^{\left(t\right)}}+{\mathbf{b}_{v\vphantom{i}}^{\left(t\right)}}
\right)\right]  
\label{eqn:update_v}
\end{flalign}

\vspace{0.5\baselineskip} \noindent
where $\mathrm{A}_{ij}$ is the adjacency value between the \textit{i}th and \textit{j}th atoms ($\delta_{ij}$), $\mathbf{W}$ ($\mathbf{W}_{s\vphantom{i}}^{\left(t\right)}\in\mathbb{R}^{{mp}\times {hp}_3}$, $\mathbf{W}_{v\vphantom{i}}^{\left(t\right)}\in\mathbb{R}^{{mp}\times {hp}_3}$) and $\mathbf{b}$ ($\mathbf{b}_{s\vphantom{i}}^{\left(t\right)}\in\mathbb{R}^{{hp}_3}$, $\mathbf{b}_{v\vphantom{i}}^{\left(t\right)}\in\mathbb{R}^{{hp}_3}$) are the learnable weight matrices and biases, respectively. The $mp$ is the dimension size of input messages, and ${hp}_3$ is the dimension size of output convoluted features.
After $N$ times of 3D graph convolutions ($N=2$ in this work), the global pooling layer takes only updated scalar features and produces a global graph embedding. The produced global embedding passes through fully-connected layers, and the outputs are concatenated to construct a final molecular feature in the 1D-array shape, which passes through a final fully-connected layer. The atom and bond features and hyperparameters used for MolNet are given in the SI. The source code is publicly available at the GitHub repository (https://github.com/CIS-group/MolNet).

\section{Conclusions}
In summary, we proposed a GNN variant, MolNet, based on the concepts that are intuitive in (organic) chemistry, for the chemical tasks. The noncovalent adjacency matrix $\mathbf{\bar{A}}$ was introduced with consideration that noncovalent spatial locations of atoms in a molecule are equally important as covalent bonds in the determination of molecular properties and interactions, exemplified by the field effect. We found that the optimal cut-off range for $\mathbf{\bar{A}}$ was 5 Å based on screening from 3 to 6 Å. The MolNet model, structured with $\mathbf{\bar{A}}$ and the weighted bond matrix $\mathbf{B}$, showed promising performance in the chemical tasks, such as classification task of protein-ligand binding and regression task of aqueous solubility, suggesting the developmental need for chemically intuitive molecular representations in DL chemistry. However, there still remain numerous issues in molecular representations and DL architectures in the chemistry domain. For example, in addition to the dynamic description of molecules including rotational variance in molecular recognition,\cite{35kim2021} the DL recognition of stereoisomers (e.g., enantiomers) is in its infancy.\cite{36pattanaik2020,37adams2021} Moreover, most convolution filters used in the conventional GNNs could be viewed as low-pass filters, smoothing the node signals across graphs. These filters would be beneficial in the task of node and edge classifications, but perhaps not in the chemistry task that heavily relies on certain atoms or functional groups in a molecule. Nonetheless, the MolNet model proposed herein promises the future construction of chemically intuitive DL models, aided by proper digital encoding of molecules.

\bibliography{achemso-demo}

\pagebreak

\section{Supporting Information}

The following contents are available free of charge as supporting information.

\vspace{0.5\baselineskip} \noindent
\underline{\textbf{CONTENTS}}

\begin{figure}[p!]
\centering
\includegraphics[width=0.5\linewidth]{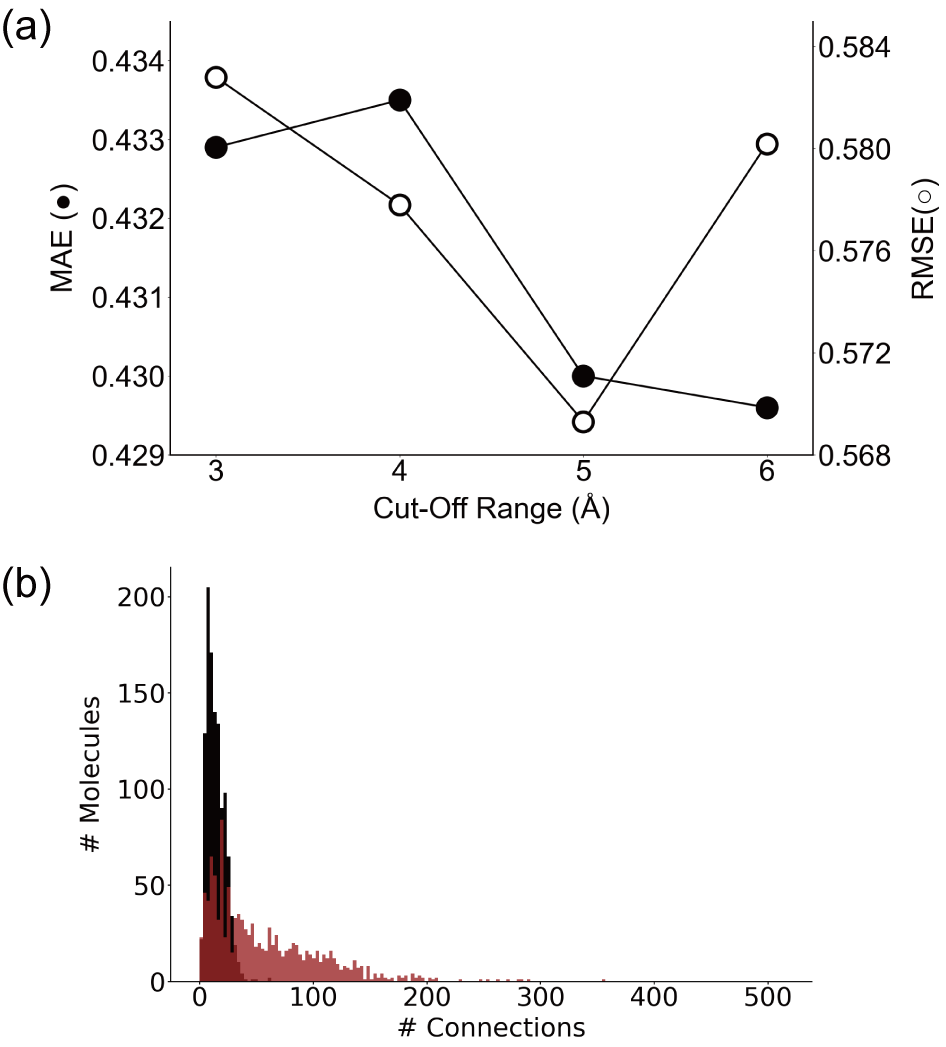}
\caption{ESOL dataset analysis. (a) Graph of AUC-ROC and AUC-PR values of versus cut-off range. (b) Histograms of (black) the number of covalent bonds and (red) the number of noncovalent interactions within 5-Å cut-off range.}
\label{fig:esol_search}
\end{figure}

\begin{itemize}
  \item MolNet details.
  \item Figure~\ref{fig:esol_search}. ESOL dataset analysis. (a) Graph of MAE and RMSE values versus cut-off range. (b) Histograms of the number of covalent bonds and the number of noncovalent interactions within 5-Å cut-off range.
  \item Table~\ref{tbl:features}. Atom and bond features.
  \item Table~\ref{tbl:hps}. Hyperparameters.
\end{itemize}

\pagebreak
\subsection{MolNet Details}

MolNet learns two 3D molecular graphs: covalent molecular graph $\mathcal{G}_C\left(\mathcal{V},\mathcal{E}\right)$ with the input of $\mathcal{B}\in\mathbb{R}^{n\times n}$, $\mathbf{X}\in\mathbb{R}^{n\times m}$, $\mathbf{E}\in\mathbb{R}^{n\times n\times e}$, and $\mathbf{R}\in\mathbb{R}^{n\times n\times 3}$, and noncovalent molecular graph $\mathcal{G}_{NC}\left(\mathcal{V},\mathcal{E}\right)$ with the input of $\mathcal{\bar{A}}\in\mathbb{R}^{n\times n}$, $\mathbf{X}\in\mathbb{R}^{n\times m}$, and $\mathbf{R}\in\mathbb{R}^{n\times n\times3}$. $\mathcal{B}$ is the normalized covalent adjacency matrix with self-loop, defined as ${\hat{\mathbf{D}}}^{-\mathbf{1}/\mathbf{2}}\hat{\mathbf{B}}{\hat{\mathbf{D}}}^{-\mathbf{1}/\mathbf{2}}$ ($\hat{\mathbf{B}}=\mathbf{B}+\mathbf{I}$, $\mathbf{D}$: diagonal node-degree (d) matrix ($\mathbf{D}$=diag(d)), $\mathbf{I}$: identity matrix), and $\mathcal{\bar{A}}$ is the normalized noncovalent adjacency matrix of $\mathbf{\bar{A}}$ with self-loop, calculated in the same way as $\mathbf{B}$. $\mathbf{X}$ is the atom-feature matrix ($n$: maximum number of atoms per molecule in the dataset, $m$: the size of total atom features), $\mathbf{E}$ is the bond-feature tensor ($n$: maximum number of bonds per molecule in the dataset, $e$: the size of total bond features), and $\mathbf{R}$ is the 3D interatom-position tensor ($\bm{r}_{ij}=\bm{p}_{j}-\bm{p}_{i}$, $\bm{p}_{i}$: 3D (\textit{x}, \textit{y}, and \textit{z})-coordinate vector of an atom \textit{i}). Atoms are represented by scalar features $\mathbf{S}_C\in\mathbb{R}^{n\times m}$ and vector features $\mathbf{V}_C\in\mathbb{R}^{n\times3\times m}$ in $\mathcal{G}_C\left(\mathcal{V},\mathcal{E}\right)$ and by scalar features $\mathbf{S}_{NC}\in\mathbb{R}^{n\times m}$ and vector features $\mathbf{V}_{NC}\in\mathbb{R}^{n\times3\times m}$ in $\mathcal{G}_{NC}\left(\mathcal{V},\mathcal{E}\right)$, where the initial scalar features are the input atom features (i.e., ${\mathbf{S}_{C}^{(0)}}\in\mathbb{R}^{n\times m}=\mathbf{X}$ and ${\mathbf{S}_{NC}^{(0)}}\in\mathbb{R}^{n\times m}=\mathbf{X}$), and the initial vector features are set to be all-zero tensor (${\mathbf{V}_{C}^{(0)}}\in\mathbb{R}^{n\times3\times m}=\mathbf{O}$ and ${\mathbf{V}_{NC}^{(0)}}\in\mathbb{R}^{n\times3\times m}=\mathbf{O}$). The 3D graph convolutions, Conv\textsubscript{C} and Conv\textsubscript{NC}, are performed independently, and the results are concatenated at the final stage. 

In the message-passing (MP) operation, a set of features ($\left\{\mathbf{S}_C,\mathbf{V}_C\right\}$ and $\left\{\mathbf{S}_{NC},\mathbf{V}_{NC}\right\}$) is updated by first mixing with the atom (and optional edge) features of neighbor atoms, leading to the generation of an intermediate atom feature. For example, the scalar-to-scalar ($s\rightarrow v$) mixing (MP$_{s\rightarrow v}$), produces $\bm{m[s\rightarrow v]}$ after nonlinear activation. The $s\rightarrow s$ and $v\rightarrow v$ mixings involve the linear combination of atom features, followed by nonlinear activation (swish for $s\rightarrow s$ and tanh for $v\rightarrow v$). The $v\rightarrow s$ interconversion is made, after linear combination of atom features of two nodes, by the multiplication of the \textit{k}-axis elements of a vector feature with ${r}_{ij,k}$ ($k$: one of the three Cartesian axes) and subsequent feature-wise summation. The $s\rightarrow v$ interconversion is done by the tensor product ($\otimes$) with $\bm{r}_{ij}$. The swish and tanh functions are used for $v\rightarrow s$ and $s\rightarrow v$ interconversions, respectively.

\begin{flalign*}
\bm{m[s\rightarrow s]}_{ij}^{\left(t\right)}=\begin{cases}
\ \left\{ 
        {
        \left (
            {
            \left ( \mathbf{S}_{i}^{\left(t\right)} \parallel \mathbf{S}_{j}^{\left(t\right)} \right)
            {\mathbf{W}_{s\rightarrow s \vphantom{i}}^{\left(t\right)}}
            +{\mathbf{b}_{s\rightarrow s \vphantom{i}}^{\left(t\right)}}
            }
            \right)} \right. \\ 
\left. \quad\quad\quad \odot
    {
    \left(\mathbf{E}_{ij}{\mathbf{W}_{e,s\rightarrow s \vphantom{i}}^{\left(t\right)}}
    +{\mathbf{b}_{e,s\rightarrow s \vphantom{i}}^{\left(t\right)}}\right)
    }\right\}
    {\mathbf{W}_{m,s\rightarrow s \vphantom{i}}^{\left(t\right)}}+{\mathbf{b}_{m,s\rightarrow s \vphantom{i}}^{\left(t\right)}}
    & \quad \text{if $\mathbf{E}$ exists} \\
\ \left(
    \mathbf{S}_{i}^{\left(t\right)} \parallel \mathbf{S}_{j}^{\left(t\right)} \right)
    {\mathbf{W}_{s\rightarrow s \vphantom{i}}^{\left(t\right)}}
    +{\mathbf{b}_{s\rightarrow s \vphantom{i}}^{\left(t\right)}} 
    & \quad \text{else}
\end{cases}
\end{flalign*}

\begin{flalign*}
\bm{m[v\rightarrow v]}_{ij}^{\left(t\right)}=\begin{cases}
\ \left\{ 
        {
        \left (
            {
            \left ( \mathbf{V}_{i}^{\left(t\right)} \parallel \mathbf{V}_{j}^{\left(t\right)} \right) {\mathbf{W}_{v\rightarrow v \vphantom{i}}^{\left(t\right)}}
            +{\tilde{\mathbf{b}}_{v\rightarrow v \vphantom{i}}^{\left(t\right)}}
            }
            \right)} \right. \\ 
\left. \quad\quad\quad \odot
    {
    \left(\mathbf{E}_{ij}{\mathbf{W}_{e,v\rightarrow v \vphantom{i}}^{\left(t\right)}} +{\tilde{\mathbf{b}}_{e,v\rightarrow v \vphantom{i}}^{\left(t\right)}}\right)
    }\right\}
    {\mathbf{W}_{m,v\rightarrow v \vphantom{i}}^{\left(t\right)}}
    +{\tilde{\mathbf{b}}_{m,v\rightarrow v \vphantom{i}}^{\left(t\right)}} 
    & \quad \text{if $\mathbf{E}$ exists} \\
\ \left(
    \mathbf{V}_{i}^{\left(t\right)} \parallel \mathbf{V}_{j}^{\left(t\right)} \right) 
    {\mathbf{W}_{v\rightarrow v \vphantom{i}}^{\left(t\right)}}
    +{\tilde{\mathbf{b}}_{v\rightarrow v \vphantom{i}}^{\left(t\right)}}
    & \quad \text{else}
\end{cases}
\end{flalign*}

\begin{flalign*}
\bm{m[v\rightarrow s]}_{ij}^{\left(t\right)}=\begin{cases}
\ \sum_{k\in\left(x,y,z\right)} {\left[
    \left\{
        {
        \left(
            {
            \left (\mathbf{V}_{i}^{\left(t\right)} \parallel  \mathbf{V}_{j}^{\left(t\right)}\right)
            {\mathbf{W}_{v\rightarrow s \vphantom{i}}^{\left(t\right)}}
            +{\tilde{\mathbf{b}}_{v\rightarrow s \vphantom{i}}^{\left(t\right)}}}
            \right)}  \right. \right.} \\ 
{\left. \left. \quad\quad\quad \odot
    {
    \left(\mathbf{E}_{ij}{\mathbf{W}_{e,v\rightarrow s \vphantom{i}}^{\left(t\right)}}
    +{\tilde{\mathbf{b}}_{e,v\rightarrow s \vphantom{i}}^{\left(t\right)}}\right)
    }\right\}
    \mathbf{W}_{m,v\rightarrow s \vphantom{i}}^{\left(t\right)}
    +{\tilde{\mathbf{b}}_{m,v\rightarrow s \vphantom{i}}^{\left(t\right)}}
    \right]_{k}} {r}_{ij,k} 
    & \quad \text{if $\mathbf{E}$ exists} \\
\ \sum_{k\in\left(x,y,z\right)} {
        \left(
            {
            \left (\mathbf{V}_{i}^{\left(t\right)} \parallel  \mathbf{V}_{j}^{\left(t\right)}\right)
            {\mathbf{W}_{v\rightarrow s \vphantom{i}}^{\left(t\right)}}
            +{\tilde{\mathbf{b}}_{v\rightarrow s \vphantom{i}}^{\left(t\right)}}}
            \right)}_{k} {r}_{ij,k} 
    & \quad \text{else}
\end{cases}
\end{flalign*}

\begin{flalign*}
\bm{m[s\rightarrow v]}_{ij}^{\left(t\right)}=\begin{cases}
\ \left[
    \left\{ 
        {
        \left (
            {\left(\mathbf{S}_{i}^{\left(t\right)} \parallel  \mathbf{S}_{j}^{\left(t\right)}\right) 
            {\mathbf{W}_{s\rightarrow v \vphantom{i}}^{\left(t\right)}}
            +{\mathbf{b}_{s\rightarrow v \vphantom{i}}^{\left(t\right)}}
            }
            \right)} \right. \right. \\ 
\left. \left. \quad\quad\quad \odot
    {
    \left(\mathbf{E}_{ij}{\mathbf{W}_{e,v\rightarrow v \vphantom{i}}^{\left(t\right)}} +{\tilde{\mathbf{b}}_{e,v\rightarrow v \vphantom{i}}^{\left(t\right)}}\right)
    }\right\}
    {\mathbf{W}_{m,v\rightarrow v \vphantom{i}}^{\left(t\right)}}
    +{\tilde{\mathbf{b}}_{m,v\rightarrow v \vphantom{i}}^{\left(t\right)}}  
    \right] \otimes \bm{r}_{ij}
    & \quad \text{if $\mathbf{E}$ exists} \\
\ \left(
    \left(
    \mathbf{S}_{i}^{\left(t\right)} \parallel  \mathbf{S}_{j}^{\left(t\right)}\right) 
    {\mathbf{W}_{s\rightarrow v \vphantom{i}}^{\left(t\right)}}
    +{\mathbf{b}_{s\rightarrow v \vphantom{i}}^{\left(t\right)}}
    \right) \otimes \bm{r}_{ij}
    & \quad \text{else}
\end{cases}
\end{flalign*}

\vspace{0.5\baselineskip} \noindent
where ${\mathbf{S}_{i}^{\left(t\right)}}$ and ${\mathbf{V}_{i}^{\left(t\right)}}$ are the scalar and vector features of the \textit{i}th atom on the layer \textit{t}, $\parallel$ represents the concatenation, 
$\mathbf{W}$ (${\mathbf{W}_{s\rightarrow s\vphantom{i}}^{\left(t\right)}}\in\mathbb{R}^{{2m}\times {hp}_1}$, ${\mathbf{W}_{v\rightarrow v\vphantom{i}}^{\left(t\right)}}\in\mathbb{R}^{{2m}\times {hp}_1}$, ${\mathbf{W}_{v\rightarrow s\vphantom{i}}^{\left(t\right)}}\in\mathbb{R}^{{2m}\times {hp}_1}$, ${\mathbf{W}_{s\rightarrow v\vphantom{i}}^{\left(t\right)}}\in\mathbb{R}^{{2m}\times {hp}_1}$),
$\mathbf{b}$ (${\mathbf{b}_{s\rightarrow s\vphantom{i}}^{\left(t\right)}}\in\mathbb{R}^{{hp}_1}$, ${\mathbf{b}_{v\rightarrow v\vphantom{i}}^{\left(t\right)}}\in\mathbb{R}^{{hp}_1}$, ${\mathbf{b}_{v\rightarrow s\vphantom{i}}^{\left(t\right)}}\in\mathbb{R}^{{hp}_1}$, ${\mathbf{b}_{s\rightarrow v\vphantom{i}}^{\left(t\right)}}\in\mathbb{R}^{{hp}_1}$),
$\mathbf{W}_{e\vphantom{i}}$ (${\mathbf{W}_{e,s\rightarrow s\vphantom{i}}^{\left(t\right)}}\in\mathbb{R}^{e\times {hp}_1}$, ${\mathbf{W}_{e,v\rightarrow v\vphantom{i}}^{\left(t\right)}}\in\mathbb{R}^{e\times {hp}_1}$, ${\mathbf{W}_{e,v\rightarrow s\vphantom{i}}^{\left(t\right)}}\in\mathbb{R}^{e\times {hp}_1}$, ${\mathbf{W}_{e,s\rightarrow v\vphantom{i}}^{\left(t\right)}}\in\mathbb{R}^{e\times {hp}_1}$),
$\mathbf{b}_{e\vphantom{i}}$ (${\mathbf{b}_{e,s\rightarrow s\vphantom{i}}^{\left(t\right)}}\in\mathbb{R}^{{hp}_1}$, ${\mathbf{b}_{e,v\rightarrow v\vphantom{i}}^{\left(t\right)}}\in\mathbb{R}^{{hp}_1}$, ${\mathbf{b}_{e,v\rightarrow s\vphantom{i}}^{\left(t\right)}}\in\mathbb{R}^{{hp}_1}$, ${\mathbf{b}_{e,s\rightarrow v\vphantom{i}}^{\left(t\right)}}\in\mathbb{R}^{{hp}_1}$),
$\mathbf{W}_{m\vphantom{i}}$ (${\mathbf{W}_{m,s\rightarrow s\vphantom{i}}^{\left(t\right)}}\in\mathbb{R}^{{2{hp}_2}\times {hp}_2}$, ${\mathbf{W}_{m,v\rightarrow v\vphantom{i}}^{\left(t\right)}}\in\mathbb{R}^{{2{hp}_2}\times {hp}_2}$, ${\mathbf{W}_{m,v\rightarrow s\vphantom{i}}^{\left(t\right)}}\in\mathbb{R}^{{2{hp}_2}\times {hp}_2}$, ${\mathbf{W}_{m,s\rightarrow v\vphantom{i}}^{\left(t\right)}}\in\mathbb{R}^{{2{hp}_2}\times {hp}_2}$) and
$\mathbf{b}_{m\vphantom{i}}$ (${\mathbf{b}_{m,s\rightarrow s\vphantom{i}}^{\left(t\right)}}\in\mathbb{R}^{{hp}_2}$, ${\mathbf{b}_{m,v\rightarrow v\vphantom{i}}^{\left(t\right)}}\in\mathbb{R}^{{hp}_2}$, ${\mathbf{b}_{m,v\rightarrow s\vphantom{i}}^{\left(t\right)}}\in\mathbb{R}^{{hp}_2}$, ${\mathbf{b}_{m,s\rightarrow v\vphantom{i}}^{\left(t\right)}}\in\mathbb{R}^{{hp}_2}$)
are the learnable weight matrices and biases, respectively. The ${hp}_1$ and ${hp}_2$ are the dimension sizes of output messages, and 
$\tilde{\mathbf{b}} \coloneqq \left[\mathbf{b}\parallel\mathbf{b}\parallel\mathbf{b}\right]^{T}$, where $\tilde{\mathbf{b}}\in\mathbb{R}^{3\times {hp}_1}$, $\tilde{\mathbf{b}}_{e\vphantom{i}}\in\mathbb{R}^{3\times {hp}_1}$, and $\tilde{\mathbf{b}}_{m\vphantom{i}}\in\mathbb{R}^{3\times {hp}_2}$. 
The same $\mathbf{W}$ and $\mathbf{b}$ are applied for all the graphs.

The graph update operation involves the aggregation of $\bm{m[s\rightarrow s]}$ and $\bm{m[v\rightarrow s]}$ for scalar features or of $\bm{m[v\rightarrow v]}$ and $\bm{m[s\rightarrow v]}$ for vector features:

\begin{flalign*}
\mathbf{S}_{i}^{\left(t+1\right)} & =\mathbf{S}_{i}^{\left(t\right)}+\mathrm{swish}\left[\mathrm{Update}_{s}\right]  \nonumber \\
&   =\mathbf{S}_{i}^{\left(t\right)}+\mathrm{swish}\left[
\sum_{j\in\left(i,\mathcal{N}_i\right)}\mathrm{A}_{ij}\left(
\left(\bm{m[s\rightarrow s]}_{ij}^{\left(t\right)}\parallel \bm{m[v\rightarrow s]}_{ij}^{\left(t\right)} \right)
{\mathbf{W}_{s\vphantom{i}}^{\left(t\right)}}+{\mathbf{b}_{s\vphantom{i}}^{\left(t\right)}}
\right)\right]  
\end{flalign*}

\begin{flalign*}
\mathbf{V}_{i}^{\left(t+1\right)}   
&   =\mathbf{V}_{i}^{\left(t\right)}+\mathrm{swish}\left[\mathrm{Update}_{s}\right]  \nonumber \\
&   =\mathbf{V}_{i}^{\left(t\right)}+\mathrm{swish}\left[
\sum_{j\in\left(i,\mathcal{N}_i\right)}\mathrm{A}_{ij}\left(
\left(\bm{m[v\rightarrow v]}_{ij}^{\left(t\right)}\parallel \bm{m[s\rightarrow v]}_{ij}^{\left(t\right)} \right)
{\mathbf{W}_{v\vphantom{i}}^{\left(t\right)}}+{\mathbf{b}_{v\vphantom{i}}^{\left(t\right)}}
\right)\right]  
\end{flalign*}

\vspace{0.5\baselineskip} \noindent
where $\mathrm{A}_{ij}$ is the adjacency value between the \textit{i}th and \textit{j}th atoms ($\delta_{ij}$), $\mathbf{W}$ ($\mathbf{W}_{s\vphantom{i}}^{\left(t\right)}\in\mathbb{R}^{{mp}\times {hp}_3}$, $\mathbf{W}_{v\vphantom{i}}^{\left(t\right)}\in\mathbb{R}^{{mp}\times {hp}_3}$) and $\mathbf{b}$ ($\mathbf{b}_{s\vphantom{i}}^{\left(t\right)}\in\mathbb{R}^{{hp}_3}$, $\mathbf{b}_{v\vphantom{i}}^{\left(t\right)}\in\mathbb{R}^{{hp}_3}$) are the learnable weight matrices and biases, respectively. The $mp$ is the dimension size of input messages and ${hp}_3$ is the dimension size of output convoluted features.
After $N$ times of 3D graph convolutions ($N=2$ in this work), the global pooling layer takes only updated scalar features and produces a global graph embedding. The produced global embedding passes through fully-connected layers, and the outputs are concatenated to construct a final molecular feature in the 1D-array shape, which passes through a final fully-connected layer.

\clearpage

\begin{table}[t]
  \caption{Atom and bond features used for molecular graphs in this work.}
  \label{tbl:features}
  {\footnotesize   
  \begin{tabular}{c|cccc}
\hline
Type & Feature               & Description                                                           & Size \\
\hline
atom & atom type             & B, C, N, O, F, Na, Si, P, S, Cl, Se, Br, Sn, I, or others (one-hot) & 15   \\
     & degree                & degree of bonded atoms (0 to 6; one-hot)                              & 7    \\
     & formal   charge       & value of formal charge (-3 to 3; one-hot)                             & 7    \\
     & implicit   valence    & value of implicit valence (0 to 6; one-hot)                           & 7    \\
     & number of   hydrogens & number of   bonded hydrogens (0 to 4; one-hot)                        & 5    \\
     & hybridization         & sp, sp$_{2}$,   sp$_{3}$, sp$_{3}$d, or sp$_{3}$d$_{2}$ (one-hot)               & 5    \\
     & aromaticity           & whether   the atom is in an aromatic system or not (binary)           & 1    \\
     & ring                  & size of   the ring the atom belonged to (ring size: 3 to 8; one-hot)  & 6    \\
     & acidity               & whether   the atom is acidic or not (binary)                          & 1    \\
     & basicity              & whether   the atom is basic or not (binary)                           & 1    \\
     & electron   donor      & whether   the atom donates electrons or not (binary)                  & 1    \\
     & electron   acceptor   & whether   the atom accepts electrons or not (binary)                  & 1    \\
\hline
bond & bond type             & single,   double, triple, or aromatic (one-hot)                       & 4    \\
     & conjugation           & whether   the bond is in conjugation or not (binary)                  & 1    \\
     & ring                  & whether   the bond is in ring or not (binary)                         & 1    \\
     & chirality             & \textit{E}, \textit{Z}, any, or none (one-hot)                        & 4    \\
\hline
  \end{tabular}
  }
\end{table}

\begin{table}[b]
  \caption{Hyperparameters used for MolNet in this work.}
  \label{tbl:hps}
  {\footnotesize
  \begin{tabular}{c|cc}
\hline
Type                       & Hyperparameter                                                & Size   \\
\hline
3D graph convolution layer & output size: scalar-to-scalar operation                       & 128    \\
                           & output size: scalar-to-vector operation                       & 128    \\
                           & output size: vector-to-scalar operation                       & 128    \\
                           & output size: vector-to-vector operation                      & 128    \\
                           & output size: scalar-feature convolution                       & 128    \\
                           & output size: vector-feature convolution                       & 128    \\
                           & layer number                                                  & 2      \\
\hline
fully connected layer      & output size                                                   & 128    \\
                           & layer number                                                  & 2      \\
                           & lambda value on L2 regularization                             & 0.005  \\
\hline
training                   & batch size                                                    & 8      \\
                           & (gradient descent method)                                     & (Adam) \\
                           & initial learning rate                                         & 0.001  \\
                           & learning-rate decay rate for CosineAnnealingDecay             & 1.0    \\
                           & initial decay step size for CosineAnnealingDecay              & 10     \\
                           & ratio of increasing multiplied value for CosineAnnealingDecay & 2.0    \\
                           & patience for early stopping                                   & 50     \\
                           & cross-validation fold                                         & 10     \\
\hline
  \end{tabular}
  }
\end{table}

\bibliographystyle{achemso}
\end{document}